\documentclass[aps,twocolumn,floatfix]{revtex4-1}

\usepackage{graphicx}

\usepackage[T1]{fontenc}       

\usepackage{amsmath}

\usepackage{lmodern}  
\usepackage{sistyle}

\global\long\def\kbo{\ensuremath{\mathbf{k}}}

\begin{document}

\title{Effect of phonon bath dimensionality on the spectral tuning of
single-photon emitters in the Purcell regime}

\author{Yannick Chassagneux}
\email{yannick.chassagneux@lpa.ens.fr}
\author{Adrien Jeantet}
\author{Th\'eo Claude}
\author{Christophe Voisin}
\affiliation{Laboratoire Pierre Aigrain, D\'epartement de physique de l'ENS, \'Ecole normale sup\'erieure, PSL Research University, Universit\'e Paris Diderot, Sorbonne Paris Cit\'e, Sorbonne Universit\'es, UPMC Univ. Paris 06, CNRS, 75005 Paris, France }

\date{\today}

\begin{abstract}
We develop a theoretical frame to investigate the spectral dependence of the brightness of a single-photon source made of a solid-state nano-emitter embedded in a high-quality factor micro-cavity. This study encompasses the cases of localized
excitons embedded in a one (1D), two (2D) or three-dimensional (3D) matrix. The population
evolution is calculated based on a spin-boson model, using the non-interacting
blip approximation (NIBA). We find that the spectral dependence of the single-photon source brightness (hereafter called
spectral efficiency) can be expressed analytically through the free-space emission and absorption spectra
of the emitter, the vacuum Rabi splitting and the loss rates of the system. In other words,
the free-space spectrum of the emitter encodes all the relevant information on the interaction between the exciton and the phonon bath
to obtain the dynamics of the cavity coupled system. We compute numerically
the spectral efficiency for several types of localized emitters differing by the phonon bath dimentionality.
In particular, in low-dimensional systems where this interaction is enhanced, a pronounced asymmetric energy exchange between the
emitter and the cavity on the phonon side-bands yields a considerable
extension of the tuning range of the source through phonon-assisted
cavity feeding, possibly surpassing that of a purely resonant system.
\end{abstract}


\maketitle


\section{Introduction}

\begin{figure}[ht]
\includegraphics[width=4.5cm]{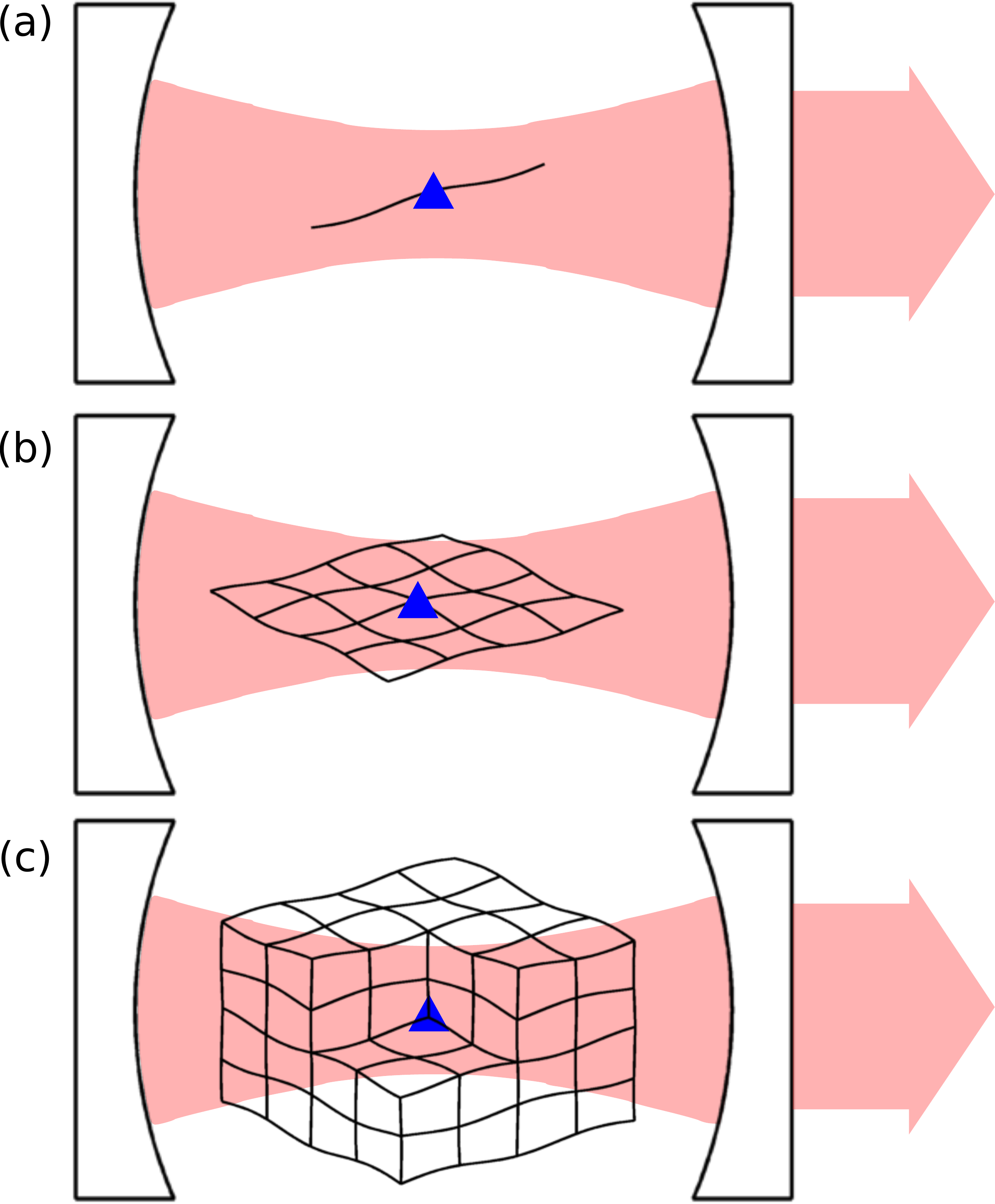} \caption{   Localized emitter coupled to a 1D (panel (a) ), 2D (panel (b)) or 3D (panel (c)) phonon bath embedded in a cavity. \label{fig:schemaillustratif}}
\end{figure}

The paradigmic situation of cavity quantum electrodynamics (CQED) consists of a two-level system interacting with a single-mode cavity. This interaction
has been widely investigated in the field of atomic physics, where
a subset of atomic levels makes up for an ideal two-level system~\cite{goy1983observation}.
However, the need for integrable devices for applications has led
to the investigation of solid-state emitters. A large variety of nano-emitters
 such as quantum-dots\cite{gerard1998enhanced} or defects in nano-diamonds~\cite{Faraon2011}
have been investigated. Depending on the system, the zero-dimensional
emitter can either be a localized exciton, or a subsets of confined
electronic levels. For the sake of simplicity, we call it exciton
throughout the paper. The move to solid-state emitters has brought additional complexity due to the vibrationnal degrees of freedom, which can turn out to be valuable to enrich the properties of the single-photon source.
 The coupling of the exciton to the phonon bath can lead from moderate to severe non-Markovian modifications of the
two-level system dynamics, which in turn leads to non-Lorentzian photoluminescence
line-shapes~\cite{PhysRevLett.87.157401,Vialla2014a,Galland2008}.
 The interaction with one or several phonon modes broadens the main line - the zero phonon line (ZPL) - through virtual processes and yields side-bands through inelastic processes. Close to the ZPL, the dominant processes imply acoustic phonons and can be described by the phonon spectral density function which scales as a power-law of the phonon energy\cite{Krummheuer2002a} $\propto(\hbar\omega)^{s}$. In the case of the deformation potential interaction the exponent $s$ is equal to the dimensionality of the phonon bath.
Three distinct regimes can be observed: super-ohmic ($s>1$), ohmic ($s=1$)
and sub-ohmic ($s<1$). For $s=3$ (or higher), the phonon coupling can be seen as a weak perturbation and the dynamics can be well captured in a perturbative approach. For $s=2$ the perturbation theory starts to be less accurate. For ohmic and sub-ohmic couplings, the exciton-phonon coupling must be treated in a non perturbative approach (polaron approach \cite{Mahan}).

The interaction of a two-level system
with a phonon bath has mainly been investigated for epitaxial quantum
dots in a three-dimensional matrix \cite{Krummheuer2002a,wilson2002quantum,roy2011influence}
(super-ohmic coupling). In recent years, the emergence of quantum
optics experiments performed with low-dimensional materials, such as
defects in transition-metal dichalcogenides~\cite{srivastava2015optically,koperski2015single,he2015single,chakraborty2015voltage},
or trapped excitons in single carbon nanotubes~\cite{Watahiki2012,JeantetPRL,JeantetNL},
has brought up new paradigms for reduced dimensionality phonon baths.

In this paper, we describe the effect of the exciton-phonon coupling
for a single 0D exciton embedded in a 1D, 2D or 3D phonon bath and
coupled to an optical microcavity. More precisely, we study the case of a localized
exciton coupled to a single acoustic phonon branch through a deformation
potential. The results can easily  be  extended to several acoustic
phonon branches, and to other types of phonon couplings such as piezoelectric
coupling (where the power law $s$ no longer scales as the bath dimensionality).

The case of the 3D phonon bath is benchmarked against the standard approximations
based on perturbations of the expectation value of the bath displacement
operator. The results in lower-dimensions are new as the usual approximations
do not hold for the 1D and 2D cases. We find that the behavior of
the exciton-phonon-cavity coupling can be traced back to the free-space
spectrum of the emitter, together with the loss rates of the emitter
and the cavity. Thus, the system dynamics is described as an incoherent exchange
of energy between two leaky boxes, one figuring the emitter and
the other the cavity. As a consequence, the full complexity of the
exciton-phonon interactions is embedded into the effective exchange
rates between the two boxes.



\section{Model}

\subsection{Hamiltonian of the system}

In this work we consider a localized exciton treated as a two-level system
described by the Pauli matrices ($\hat{\sigma}^{+}$ , $\hat{\sigma}^{-}$).
More precisely, we consider the following Hamiltonian: 
\begin{equation}
\hat{H}=\hat{H}_{0}+\hat{H}_{X-p}+\hat{H}_{X-cav},
\end{equation}
where $H_{0}$ is the non-interacting Hamiltonian which reads: 
\begin{equation}
\hat{H}_{0}=\hbar\omega_{X}\hat{\sigma}^{+}\hat{\sigma}^{-}+\hbar\sum_{\kbo}\omega_{\kbo}\hat{b}_{\kbo}^{\dag}\hat{b}_{\kbo}+\hbar\omega_{cav}\hat{a}^{\dag}\hat{a},
\end{equation}
where $\hbar\omega_{X}$ , $\hbar\omega_{\kbo}$ and $\hbar\omega_{cav}$
are respectively the energies of the exciton, of the phonon mode with
wavevector $\kbo$ and of the cavity mode. $\hat{b}_{\kbo}^{\dag}$
(resp. $\hat{b}_{\kbo}$) is the creation (resp. annihilation)
operator of a phonon with wavevector $\kbo$. $\hat{a}$ and $\hat{a}^{\dag}$
are the usual cavity mode operators. The exciton-phonon coupling is
given by: 
\begin{equation}
\hat{H}_{X-p}=\hbar\hat{\sigma}^{+}\hat{\sigma}^{-}\sum_{\kbo}\left(g_{\kbo}\hat{b}_{\kbo}+g_{\kbo}^{*}\hat{b}_{\kbo}^{\dag}\right).
\end{equation}
The expression of the coupling parameter $g_{\kbo}$ depends on the
interaction mechanism, and can be described in the continuous limit
by the exciton-phonon spectral density function $J$\cite{Leggett1987}, defined
in this work as: $J(\omega)\equiv\pi\sum_{\kbo}|g_{\kbo}|^{2}\delta(\omega-\omega_{\kbo})$.
In the case of a localized exciton coupled through the deformation potential
$D$ to acoustic phonons (with linear dispersion $\omega_{\kbo}=v|\kbo|$),
the spectral density reads: 
\begin{equation}
J(\omega)=\frac{2\pi\alpha}{\omega_c^{s-1}} \ \omega^s\  |\mathcal{F}(\omega)|^{2}.
\end{equation}
For the sake of simplicity, the exponent $s$ will be called the \textit{bath
dimensionality} in the rest of the paper. $\mathcal{F}$ is related
to the momentum conservation and depends on the enveloppe wavefunction of the
localized exciton. For a Gaussian enveloppe, it reads~\cite{Krummheuer2002a}
$\mathcal{F}=\exp(-(\omega/\omega_{c})^{2})$, where the energy cut-off
$\omega_{c}$ is related to the exciton localization length $\sigma$
by $\omega_{c}=2v/\sigma$, where $v$ is the sound velocity. In low dimensional systems, modifications in the phonon amplitude and density of states (as for instance the creation of a phonon gap through mechanical interaction with the surrounding \cite{Vialla2014a}) can be taken into account through a modified form of $|\mathcal{F}(\omega)|$. With these modifications, if $\mathcal{F}(\omega)$ tends to zero for a vanishing phonon energy, an effective  exponent $s$ can be defined at low energy, yielding an effective bath dimensionality higher than the physical dimensionality and caracteristic signatures in the luminescence spectrum of the emitter \cite{Vialla2014a}. 
Conversely, the choice of the cut-off function (for large $\omega$) has a limited impact on the results of our study: the optical spectra would be slightly modified, but the comparison of the emission in cavity to the one in free-space would be hardly modified.

The dimensionless parameter $\alpha$ is given by: $\alpha=\frac{D^{2}}{4\pi\hbar v^{3}\rho}\frac{\sqrt{\pi}}{\Gamma[s/2]}\left(\frac{\omega_{c}}{2\sqrt{\pi}v}\right)^{s-1}$
where $\rho$ is the volumic (resp. linear, areal) mass density
in 3D (resp. 1D, 2D), and $\Gamma$ is the gamma function. This
dimensionless parameter can be determined experimentally or derived from the parameters of the material. It can vary slightly,
from $\alpha=0.17$, with $\hbar\omega_{c}\approx \SI{1}{meV}$ in
the case of quantum dots (see  Portalupi et al.~\cite{portalupi2015bright} )
to $\alpha=0.29$ with $\hbar\omega_{c}\approx \SI{9}{meV}$ in the
case of nanotubes (see Galland et al.~\cite{Galland2008}).

The exciton-cavity mode coupling  is considered within the Jaynes-Cummings approximation: 
\begin{equation}
\hat{H}_{X-cav}=i\hbar g\left(\hat{a}^{\dag}\hat{\sigma}^{-}-\hat{a}\hat{\sigma}^{+}\right),
\end{equation}
where $2g$ is the vacuum Rabi splitting.

Without the cavity, the model is equivalent to the so-called \textit{independent
boson model}~\cite{Krummheuer2002a} which can be diagonalized exactly.
In the presence of the cavity, and as long as a single excitation
is considered, the model can be mapped onto the \textit{spin boson
model}~\cite{Leggett1987}, in which the two values of the spin would
correspond to the presence of an elementary excitation either in the
cavity mode or in the exciton. Following the work of Leggett et al.~\cite{Leggett1987},
we will use the so-called \textit{non-interacting blip approximation}
(NIBA) to obtain the reduced dynamics, \textit{i.e.} the
population evolution of the exciton and of the cavity mode.

\subsection{Free-space emission and absorption spectra}
By using a polaron transformation, the freespace emission and absorption spectra can be obtained without any approximation \cite{Krummheuer2002a,Mahan} and are given by (see Appendix \ref{annexefs} for the derivation):
\begin{align}
S^{abs}(\omega) & = & 2\int_{0}^{\infty}\text{Re}\Big[e^{i(\omega-\overline{\omega}_{X})t-\frac{\gamma+\gamma^{*}}{2}t}K(t)\Big]dt,\nonumber\\
S^{emi}(\omega) & = & 2\int_{0}^{\infty}\text{Re}\Big[e^{i(\omega-\overline{\omega}_{X})t-\frac{\gamma+\gamma^{*}}{2}t}K^{*}(t)\Big]dt,\label{eq:fsspectra}
\end{align}
where $\hbar \overline{\omega}_X$ is the transition energy (renormalized by the polaron shift). $\gamma$ is the total recombination rate (radiative and non radiative) of the emitter and $\gamma^*$ is its pure dephasing rate. The factor two before the integral ensures the following normalization condition $\int S(\omega)d\omega=2\pi$. 
The phonon kernel $K(t)$ can be computed exactly, and reads in the
continuous limit~\cite{Leggett1987}: 
\begin{eqnarray}
& K(t) & =  \exp\Big(-i\frac{Q_{1}(t)}{\pi}-\frac{Q_{2}(t)}{\pi}\Big),\\
\text{with}\ \  & Q_{1}(t) & =  \int_{0}^{\infty}\frac{J(\omega)}{\omega^{2}}\sin(\omega t)d\omega,\nonumber\\
\text{and} \ \ & Q_{2}(t) & =  \int_{0}^{\infty}\frac{J(\omega)}{\omega^{2}}\big[1-\cos(\omega t)\big]\text{coth}\Big(\frac{\hbar\omega}{2k_{b}T}\Big)d\omega,\nonumber \label{eqKQ1Q2}
\end{eqnarray}
where $k_{b}$ is the Boltzmann constant and $T$ the temperature. $Q_1$ and $Q_2$ represent the phase and damping contributions to the phonon kernel, respectively.


\begin{figure}[ht]
\includegraphics[width=8.6cm]{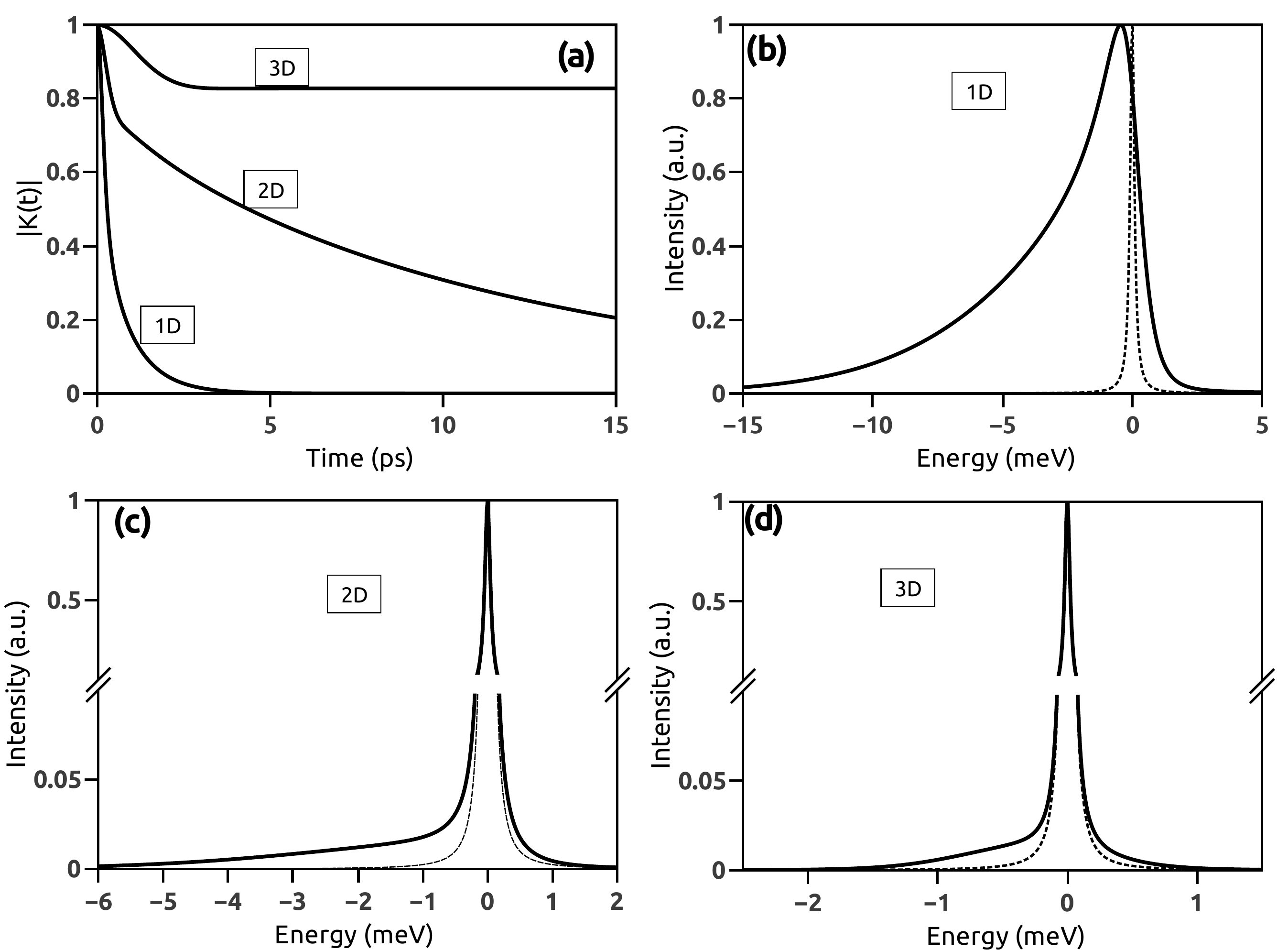} \caption{(a) Time evolution of the modulus of the phonon kernel function $|K(t)|$
for different bath dimensionalities. The temperature is set to 5K. The
parameters are $\alpha=0.29$, $\hbar\omega_{c}=\SI{8.9}{meV}$ in
1D~\cite{Galland2008}, $\alpha=0.2$, $\hbar\omega_{c}=\SI{5}{meV}$
in 2D, and $\alpha=0.17$, $\hbar\omega_{c}=\SI{1.09}{meV}$
in 3D~\cite{portalupi2015bright} (see main text for details). (b) (c), and (d) are the corresponding
emission spectra in 1D, 2D and 3D calculated with eq.~\eqref{eq:fsspectra},
and using $\gamma+\gamma^{*}=\SI{200}{\micro eV}$ in 1D, $\gamma+\gamma^{*}=\SI{100}{\micro eV}$
in 2D, and $\gamma+\gamma^{*}=\SI{50}{\micro eV}$ in 3D. The absorption
spectra can be deduced from the emission one through a mirror symmetry with respect to the
ZPL position (set to 0~eV in these plots). The dashed lines in panel (c) and (d) correspond to the emission spectra in the absence of phonon coupling.\label{fig:spectre}}
\end{figure}

The shape of the spectra is mainly determined by the behaviour of the phonon spectral density $J(\omega)$ near the origin, which is directly related to the dimensionality of the phonon bath.
The modulus of the phonon kernel $|K(t)|$ is plotted in fig.~\ref{fig:spectre}a
in three different cases at a temperature of 5~K. In the 1D case, we use $\alpha=0.29$, and $\hbar\omega_{c}=\SI{8.9}{meV}$ which
correspond to a localized exciton in a carbon nanotube, studied
by Galland et al.~\cite{Galland2008}. The corresponding free-space emission
spectrum, obtained from eq.~\eqref{eq:fsspectra} is shown in fig.~\ref{fig:spectre}b,
for $\hbar(\gamma+\gamma^{*})=\SI{200}{\micro eV}$~\cite{Galland2008}.
We note that the pure dephasing contribution has a limited impact on the line profile as the coupling
to phonons yields a much stronger broadening of the ZPL (proportional
to the temperature, see Appendix \ref{sec:valNIBA} for details). In the 2D case, to the best of our
knowledge, $\alpha$ and $\omega_{c}$ have not been experimentally
measured yet. We use $\alpha=0.2$ and $\hbar\omega_{c}=\SI{5}{meV}$
which would correspond to an exciton localized over $\SI{1.5}{nm}$
in a transition metal dichalcogenide layer (the sound velocity can be found
in ref.~\cite{zhang2014two}). The corresponding free-space spectrum
in fig.~\ref{fig:spectre}c is plotted for $\gamma+\gamma^{*}=\SI{100}{\micro eV}$
(linewidth measured by Srivastana et al.~\cite{srivastava2015optically}).
For the 3D case, $\alpha=0.17$, $\hbar\omega_{c}=\SI{1.09}{meV}$
correspond to the InGaAs quantum dot investigated by Portalupi
et al. \cite{portalupi2015bright}, and the dephasing term is $\gamma+\gamma^{*}=\SI{50}{\micro eV}$
for the free-space spectrum of fig.~\ref{fig:spectre}d.

The behaviour of the phonon kernel varies strongly with the phonon
bath dimensionality. In 1D, the decay is close to an exponential,
with a rate proportional to the temperature (cf Appendix \ref{sec:Klongtime}). In 2D,
after a fast initial decay, $|K(t)|$ follows a power law.
And finally in 3D, the phonon Kernel quickly tends to a constant value close to one at low temperature
and smaller at higher temperature.
The emission and absorption spectra are the convolution of the Fourier
transform (see eq.~\eqref{eq:fsspectra}) of the phonon kernel $K(t)$
and of a Lorenztian (due to the dephasing). The emission profile at low energy (in the vicinity of the ZPL) is determined by the long-time
behaviour of $K(t)$. If $K(t)$ tends to a constant value, which is the
case of the 3D phonon bath, a sharp zero-phonon line (ZPL) is obtained with
a linewidth given solely by the dephasing rate. The coupling to the phonon bath only results in tiny phonon wings. For a 2D bath, the emission spectrum
has an almost Lorentzian shape since pure dephasing usually results in a stronger broadening than the Fourier Transform of $K(t)$. Again, the coupling to the phonon bath results in weaker phonon wings. In contrast, in the 1D case the absence of any long-lasting tail in $|K(t)|$ results in
the suppression of the ZPL which is completely merged with the phonon
wings. In all cases, the intensity of the red and blue phonon side bands is
asymmetric due to the suppression of phonon absorption (responsible for the blue side-band) at
low temperature. From eq.~\eqref{eq:fsspectra}, the absorption spectra
can be deduced by a mirror symmetry around the polaron energy (which
is set to zero in fig.~\ref{fig:spectre}).

\subsection{Population evolution in the presence of the cavity}

In order to evaluate the efficiency of the coupling to the cavity mode, we first derive the dynamics of the population in the cavity-exciton system. The population evolution is described by the following set of equations :
\begin{align}
\frac{d\langle\hat{a}^{\dag}\hat{a}\rangle}{dt} & =  -\kappa\langle\hat{a}^{\dag}\hat{a}\rangle+F(t),\nonumber \\
\frac{d\langle\hat{\sigma}^{+}\hat{\sigma}^{-}\rangle}{dt} & =  -\gamma\langle\hat{\sigma}^{+}\hat{\sigma}^{-}\rangle-F(t),\label{eq:PopEvolu1}
\end{align}
where $\kappa$ is the cavity decay rate.
A full derivation of the $F(t)$ term can be found in the Annexe \ref{sec:cavcoupl}. Importantly, the expression of $F(t)$ brings  some composite terms (composed of phonon and cavity or exciton populations), which yields a non closed set of equations.
To solve this problem we make use of the so-called Non Interacting Blip Approximation (NIBA) \cite{Leggett1987,Dekker1987}. Briefly, the NIBA consists in decoupling composite terms into products of exciton (or cavity) observables with phonon observables. This approximation can be used in the weak cavity coupling regime (details on the NIBA and its validity can be found in the Annexe \ref{sec:cavcoupl}). 

Let's now focus on the single-photon generation. To this end, we compute the dynamics of the system once a single excitation (either a photon or an exciton) is launched in the system. Under this condition $F(t)$ reads:
\begin{align}
  F(t)&=2g^{2}\text{Re}\Bigg[\int_{0}^{t}ds\ e^{(i\delta-\frac{\gamma_{all}}{2})(t-s)} \nonumber \\
  &\times\Big(K^{*}(t-s)\langle\hat{\sigma}^{+}\hat{\sigma}^{-}(s)\rangle
   -K(t-s)\langle\hat{a}^{\dag}\hat{a}(s)\rangle\Big)\Bigg].\label{eq:Population_Evolution_apres_NIBA}
\end{align}
Note that for practical puposes in interpreting experimental data, equations \eqref{eq:PopEvolu1} and \eqref{eq:Population_Evolution_apres_NIBA} can possibly be computed on a purely experimental ground using parameters that are all accessible to measurements.

\section{Results}

\subsection{Single-photon spectral efficiency}

A decisive feature for single-photon sources is their tuning range,
\textit{i.e.} the spectral range on which they display a sizable brightness.
In this section, we investigate this aspect by deriving the single-photon
source efficiency as a function of the cavity detuning and we show
that the working range of the source is widely enhanced by the coupling
to a phonon bath. 

In order to obtain the spectral efficiency, we consider that one quantum
of energy $\hbar\omega_{X}$ is launched in the emitter at $t=0$
and subsequently leaks out via the cavity losses $\kappa$. This corresponds
to the  experimental situation of a pulsed off-resonance excitation. The decay can involve
several incoherent exchange cycles between the cavity and the emitter.
The brightness of the single-photon source, is thus given by the integrated
probability of decay in the cavity mode~\cite{hornecker2016influence},
$\beta=\kappa\int_{0}^{\infty}dt\langle a^{\dagger}a\rangle$.

At this stage, it is convenient
to introduce $\tilde{S}^{emi}$ and $\tilde{S}^{abs}$ defined as the emission or absorption spectra convoluted with the
Lorentzian cavity profile: $\tilde{S}^{abs/emi}(\omega)\equiv\big(S^{abs/emi}*L_{cav}\big)(\omega)$, where $L_{cav}(\omega)=\frac{2}{\pi\kappa}\frac{1}{1+\left(2\omega/\kappa\right)^{2}}$  ensures the normalisation $\int d\omega\tilde{S}^{abs/emi}(\omega)=2\pi$. 
%
$\tilde{S}^{emi}$ and $\tilde{S}^{abs}$ can either be obtained from the experimental measurement of the cavity and emitter spectra or they can be calculated on a theoretical ground using equation \eqref{eq:fsspectra} provided that the dephasing term $\frac{\gamma+\gamma^{*}}{2}$ is replaced by the total dephasing
term including the cavity contribution \textit{i.e.} $\frac{\gamma+\gamma^{*}+\kappa}{2}$.
To avoid possible confusion, we stress that $\tilde{S}^{emi}(\omega)$ is not the output emission spectra of the emitter coupled to the cavity. In fact, in the  weak coupling regime this latter is proportional to the standard product $S^{emi}(\omega) \times L_{cav}(\omega-\omega_{cav})$ \cite{Auffeves2009}.

By integrating eq. \eqref{eq:PopEvolu1} and \eqref{eq:Population_Evolution_apres_NIBA}
between $0$ and $\infty$, with the initial condition $\langle \hat{\sigma}^+ \hat{\sigma}^- \rangle_{t=0}=1$ and $\langle \hat{a}^\dag a\rangle_{t=0}=0$,
one obtains an analytical expression for the single-photon source brightness for a cavity tuned at $\omega_{cav}$: 
\begin{equation}
\beta(\omega_{cav})=\frac{g^{2}\tilde{S}^{emi}(\omega_{cav})/\gamma}{1+g^{2}\tilde{S}^{emi}(\omega_{cav})/\gamma+g^{2}\tilde{S}^{abs}(\omega_{cav})/\kappa}.\label{eq:formule_beta}
\end{equation}

\begin{figure}[ht]
\includegraphics[width=8.6cm]{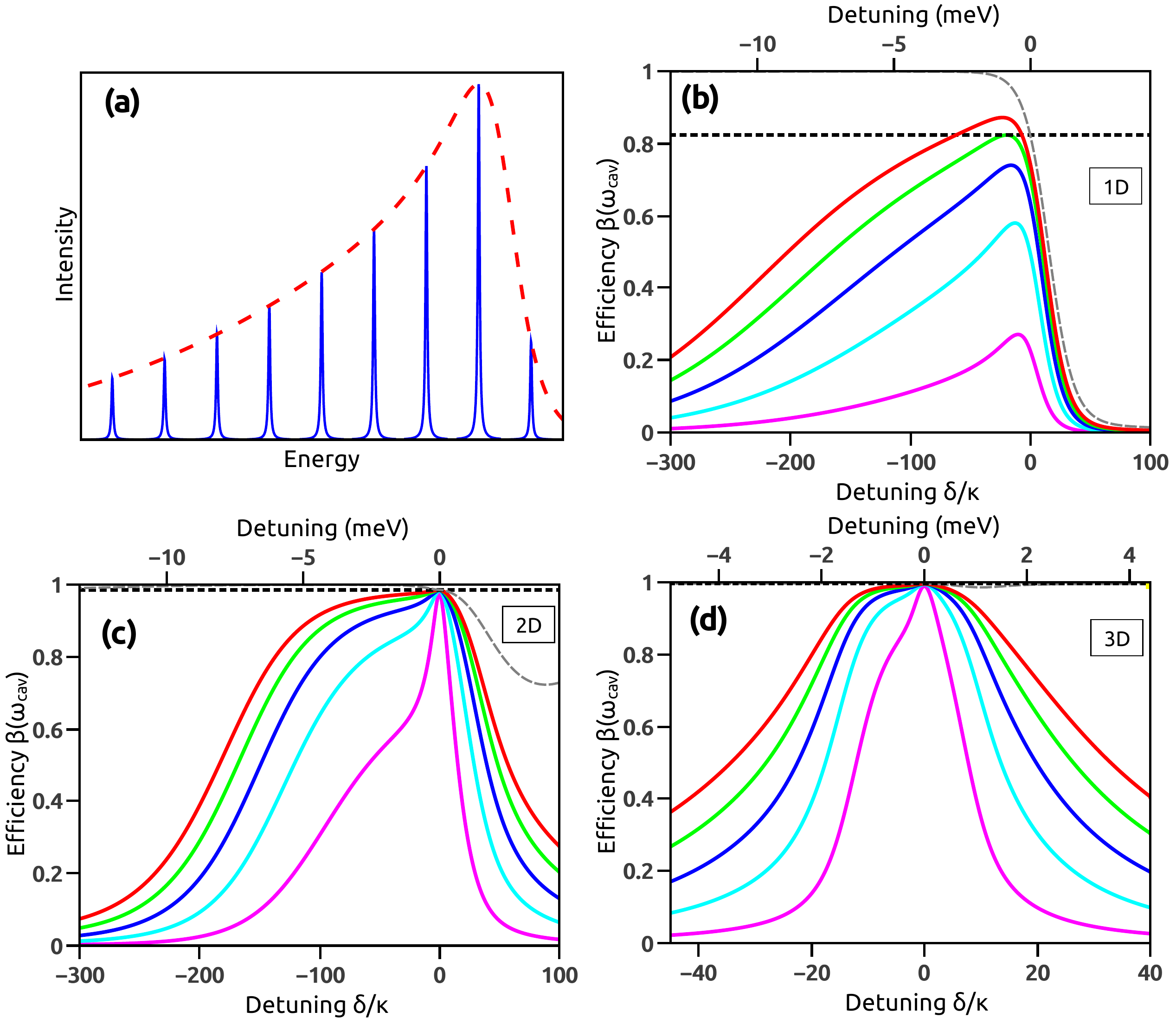} \caption{ 
(a) Schematic output spectra for different detuning. The envelope (red dashed line) is proportional to the single-photon spectral efficiency $\beta(\omega)$. (b), (c) and (d) are the single-photon spectral efficiency as a function of the cavity detuning
for different bath dimensionalities at $\SI{5}{K}$. Panels (b),
(c) and (d) correspond respectively to the 1D, 2D and 3D cases presented
in figure \ref{fig:spectre}. In each panel, the five plots correspond
to a coupling strength $\hbar g=$ 50, 100, 150, 200 and $\SI{250}{\micro eV}$.
The horizontal dashed black line is the resonant limit given by $\kappa/(\kappa+\gamma)$
(see main text). The gray dashed line is the asymptotic behavior at large coupling $g$. We use a cavity lifetime of $1/\kappa=\SI{15}{ps}$ for
the 1D~\cite{JeantetPRL} and 2D cases, and $\SI{6}{ps}$
in 3D~\cite{portalupi2015bright}. The lifetime $1/\gamma$ is $\SI{70}{ps}$ in 1D~\cite{JeantetPRL}, $\SI{1}{ns}$ in 2D~\cite{koperski2015single,he2015single,chakraborty2015voltage,srivastava2015optically},
and $\SI{1.3}{ns}$ in 3D~\cite{portalupi2015bright}.    \label{fig:beta}}
\end{figure}

Remarkably in this expression, all the complexity of the exciton-phonon interaction is encoded in the free-space absorption and emission spectra resulting in a simple analytical formula for the global output of the tripartite exciton-phonon-cavity system.
Thus, this expression can be used in two ways. From an experimental point of view it brings a means to compute the spectral efficiency of the source from the experimentally measured free-space emission spectrum, a situation of common practical interest (the absorption
spectrum $S^{abs}$ can be deduced by a mirror symmetry of the luminescence spectrum with respect to the ZPL position \cite{Mahan,Krummheuer2002a}). 
For instance, in the case of NV centers in diamonds, where the phonon wings are
mainly caused by optical phonons, Albrecht et al.~\cite{Albrecht2013}
have decomposed the Purcell effect in a heuristic sum over all vibronic transitions ending up with an expression that can
be equivalently put in the form of eq.~\eqref{eq:formule_beta}. Alternatively, this expression can be used to gain a deeper insight into the miscroscopic foundations of the single-photon source efficiency by connecting it to the electron-phonon interaction through eqs. \eqref{eq:fsspectra} and \eqref{eqKQ1Q2}.

%
%

The only free parameter of this model is the strength of the exciton-cavity
coupling $g$. As a consequence, expression~\eqref{eq:formule_beta}
can be used to predict the single-photon source efficiency of an emitter
for a given cavity design. Conversely, the measurement of the single-photon
spectral efficiency of an emitter 
can be fitted to eq.~\eqref{eq:formule_beta} in order to retrieve
the effective coupling strength (including \textit{e.g.} the spatial or polarization mode matching) \cite{JeantetNL}.

Let's note that in the absence of phonon coupling, the emission and absorption spectra are identical (Lorentzian shape) and the expression~\eqref{eq:formule_beta} of $\beta$ becomes similar to the
one obtained  by Auffeves et al. \cite{Auffeves2010}. In that latter case, it remains valid in the strong coupling regime.

The spectral efficiency of the single-photon
source is plotted in fig.~\ref{fig:beta} as a function of the detuning $\delta$ expressed
in units of the cavity linewidth $\kappa$. The parameters $\alpha$ and
$\hbar\omega_{c}$ are the same as in fig.~\ref{fig:spectre}. As
the coupling $g$ between the emitter and the cavity is increased,
the tuning range of the single-photon source is increased, eventually
reaching a hundred times $\kappa$ for realistic values of $g$ in
1D and 2D. In the 3D case, the single-photon efficiency is broader
than the cavity (and the emitter's) linewidth, but remains peaked
around the ZPL \cite{portalupi2015bright}.

The comparison of the figures~\ref{fig:beta} (b), (c) and (d) calls two major comments. First, the absolute value of the efficiency peaks to almost 1 for any coupling strength $g$ in the 2D and 3D cases, whereas it rapidely decreases for low $g$ values in 1D. This essentially arises from the choice of parameters made to describe the emitter properties, with an almost perfect free-space luminescence quantum yield in 2D and 3D (in agreement with experimental systems such as defects in TMDs or epitaxial QDs). In contrast, the 1D case is inspired from carbon nanotubes which show consistently luminescence quantum yields of the order of a few percent only \cite{JeantetPRL}. As a consequence, the main outcome of the Purcell effect for such dim emitters is a strong gain in the overall quantum yield, that roughly scales like the Purcell factor $F_p$ in the low coupling limit. As a matter of fact, it was shown experimentally that a carbon nanotube with an intrinsic quantum yield of 2\% can be brought to an effective brightness of up to 30\% when coupled to an appropriate fibered cavity reaching $g\simeq$\SI{50}{\micro eV} \cite{JeantetNL} and almost 50\% with a plasmonic resonator \cite{Luo2017}. The second qualitative observation regards the tuning range of the single-photon source. In 1D and 2D, it spans typically several hundreds of cavity linewidths whereas in the 3D case it is limited to a few tens of cavity linewidths. This is indeed the main outcome of the exciton-phonon coupling in lower dimensional systems : the strength of the effective exciton phonon coupling at low phonon frequency allows for an efficient cavity feeding effect far beyond spectral filtering. The emitter is forced by the cavity far from its intrinsic resonance but it still emits photons with a brightness close to its peak value. Finally, for ``bad'' emitters, the brightness of the source can even become larger at red detunings than at strict resonance as can be observed in fig.\ref{fig:beta}~(b). In this situation the asymmetry between the phonon-assisted absorption and emission processes at low temperature allows for an energy recycling between the exciton and the cavity even in an incoherent regime.

It is enlightning to discuss the asymptotic behavior of the efficiency
in the limit of a high cavity coupling $g$, that is for $g$ larger than the total dephasing rate. For large detunings,
the system can still be described in the adiabatic approximation as long as $\delta\gg g$.
Near the resonance, the model is no longer valid when the strong coupling
regime is reached since the NIBA cannot be used anymore. However, the model still holds in the absence of phonons~\cite{Auffeves2010}.
Since the weight of the phonon contribution near the ZPL is limited in the 3D case (as visible in figure
\ref{fig:spectre}d), we can gain an acceptable approximation
of the single-photon efficiency by taking $g \rightarrow \infty$ 
in eq. \eqref{eq:formule_beta}. For
a Lorentzian line-shape with identical emission and absorption spectra,
this limit reads 
\begin{equation}
 \beta_{g\rightarrow \infty (\delta=0)} =  \frac{\kappa}{\kappa+\gamma},
 \label{eq:peak_efficiency}
\end{equation}
 (photonic losses
over total losses). This means that the maximum efficiency is limited by the
presence of non-coherent re-absorption processes. In the presence of
phonon coupling, the picture is more subtle : at low temperature,
the absorption of phonons is strongly suppressed, which induces an asymmetry in the optical re-absorption process, depending on
the sign of the detuning. In 3D however, the magnitude of these phonon wings remains comparable to the tail of the Lorentzian ZPL and the phonon induced enhancement of the source brightness is barely observable.
In contrast, in 2D and most notably in 1D, this effect is dramatic (see dashed line in Fig.\ref{fig:beta} (c) and (d). Therefore, for such lower dimension nano-emitters a red detuned cavity offers an interesting means to overcome the natural efficiency limit $\kappa/(\kappa+\gamma)$. 
In 1D, the asymptotic limit of the spectral efficiency
 is a thermally broadened step function (though the behavior near the resonance cannot be precisely inferred from our model) reflecting the ratio of the phonon assisted absorption and emission processes.

\begin{figure}[ht]
\includegraphics[width=8.6cm]{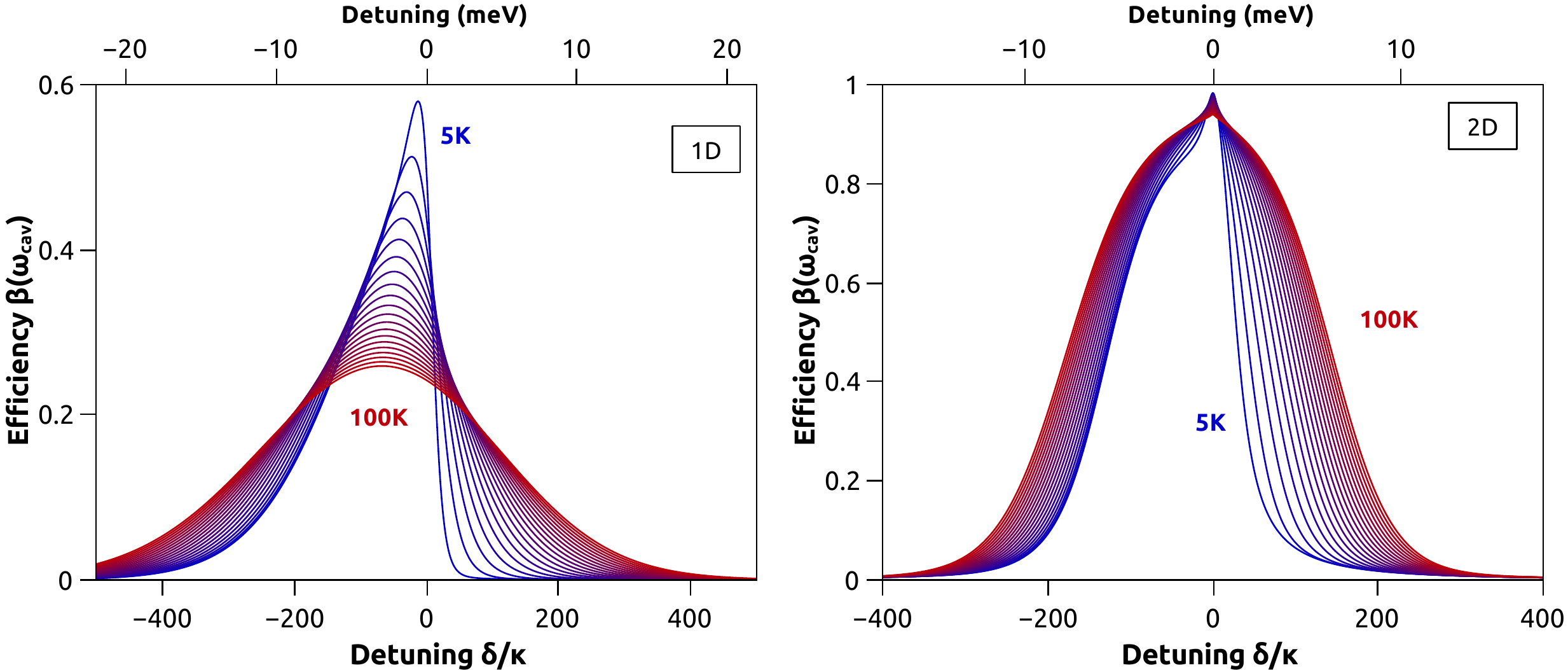} \caption{Evolution of the single-photon source spectral efficiency as a function of the detuning for several temperatures between 5 and 100 K in the 1D (a) and 2D (b) cases calculated for $g=\SI{100}{\micro eV}$. }
\label{fig:temperature}
\end{figure}

Interestingly, the effect of the temperature can readily be investigated within our model. Essentially, the increase of the phonon occupation number yields a symmetrization of the phonon emission and absorption processes and therefore a symmetrization of the spectral efficiency for red and blue detunings. We focus on the 1D and 2D cases that are more relevant experimentally because emitters such as carbon nanotubes or defects in transition metal dichalcogenides still display a bright emission at higher temperature in contrast to their 3D counterparts such as self-assembled quantum dots. As shown in Fig.~\ref{fig:temperature}, at higher temperature the spectral efficiency is basically enhanced on the blue side band and to a lower extend on the red wing leading to a symmetrization and broadening of the profile. In 1D, the peak efficiency is lower than in 2D, essentially because of the respective values of $\gamma$ that reflect that of practical emitters : in fact the strong nonradiative component of the decay in carbon nanotubes yields a larger $\gamma$ ($1/\gamma_{1D}=\SI{70}{ps} \ll 1/\gamma_{2D}=\SI{1}{ns}$) that in turn yields a lower peak efficiency (see eq. \eqref{eq:peak_efficiency}). Similarly, the weaker decrease of this peak value with the temperature is related to the initial value of $\gamma$ : for a coupling of $g=$\SI{100}{\micro eV} the 2D emitter reaches a peak efficiency close to 1 ($\gamma \ll \kappa$) and is therefore quite immune to a spreading of the oscillator strength into phonon assisted transitions (that can be seen as an effective increase of $\gamma$). In contrast, for the 1D emitter $\gamma \simeq \kappa$ even in the low temperature regime and any increase in $\gamma$ immediately translates into a decrease of the peak efficiency.

\subsection{Markovian approximation for population dynamics}

Although brightness is the most straightforward quantity to assess the technological assets of a single-photon source, time-resolved data, as obtained for instance through time-resolved photo-luminescence measurements, are valuable to gain a more direct and insightfull knowledge on the dynamics of the system. Such data can be compared to the outcome of our model through the average population $\langle\hat{a}^{\dag}\hat{a}\rangle(t)$ obtained from
equations \eqref{eq:PopEvolu1} and \eqref{eq:Population_Evolution_apres_NIBA}. 
In principle these equations can be solved numerically, nevertheless they remain
non-local in time. Here we show that one can further simplify the description of the population dynamics through an adiabatic elimination of the coherent effects in the phonon bath and in the exciton-cavity coupling. This can be done by replacing  $\langle\sigma^{+}\sigma^{-}(s)\rangle$
and $\langle a^{\dag}a(s)\rangle$ by their values
at time $t$ in eq.~\eqref{eq:Population_Evolution_apres_NIBA} and by letting the integral limit tend to infinity (owing to the exponential decay imposed by the dephasing processes). One finally obtains a (much simpler) set of equations
for the population evolution in the Markovian approximation : 
\begin{align}
   & \frac{d\langle\hat{\sigma}^{+}\hat{\sigma}^{-}\rangle}{dt}=-(\gamma+g^{2}\tilde{S}_{\omega_{cav}}^{emi})\langle\hat{\sigma}^{+}\hat{\sigma}^{-}\rangle+g^{2}\tilde{S}_{\omega_{cav}}^{abs}\langle\hat{a}^{\dag}\hat{a}\rangle\nonumber ,\\
   & \frac{d\langle\hat{a}^{\dag}\hat{a}\rangle}{dt}=-(\kappa+g^{2}\tilde{S}_{\omega_{cav}}^{abs})\langle\hat{a}^{\dag}\hat{a}\rangle+g^{2}\tilde{S}_{\omega_{cav}}^{emi}\langle\hat{\sigma}^{+}\hat{\sigma}^{-}\rangle,\label{markovdynamics}
\end{align}
where the emission and absorption probabilities $\tilde{S}$
are taken at the cavity frequency $\omega_{cav}$. Using this adiabatic elimination,
the population evolution becomes Markovian but the non-Markovian contribution
to the decoherence remains encoded in $\tilde{S}$ and is responsible for the typical asymmetric spectra. Let's emphasize
that in this approximation, the expression of the single-photon
efficiency $\beta$ does not differ from eq.~\eqref{eq:formule_beta}.
In other words, a fast modulation of the population can possibly
be lost, but the time-integrated values remain accurate. 

\begin{figure}[ht]
\includegraphics[width=8.6cm]{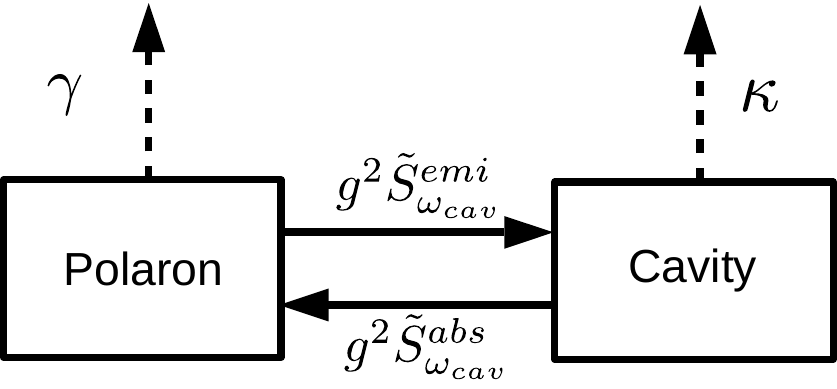} \caption{Sketch of the system as a set of two coupled boxes representing the emitter and the cavity respectively . The exchanges between these two boxes are
governed by the coupling rate $g$ and the free-space absorption/emission
probability $\tilde{S}_{\omega_{cav}}^{abs}$/$\tilde{S}_{\omega_{cav}}^{emi}$, taken at the cavity
 frequency. The emitter can also decay towards the environment
at a rate $\gamma$, while a photon in the cavity can leak out at
a rate $\kappa$. \label{fig:twoboxes}}
\end{figure}

In this adiabatic approximation, the system can schematically be depicted as a pair of boxes exchanging
energy, with losses towards the environment, as shown in fig.~\ref{fig:twoboxes}.
The two boxes represent the polaron (exciton plus phonons) and the cavity
mode, with loss rates $\gamma$ and $\kappa$ respectively. A quantum
of energy can go from the polaron box to the cavity box through the
emission of a photon. This process is proportional to
the emission probability $\tilde{S}_{\omega_{cav}}^{emi}$. Conversely,
the absorption of a photon, with a probability proportional to $\tilde{S}_{\omega_{cav}}^{abs}$
brings a quantum of energy from the cavity box to the polaron box.
Note that this energy exchange is asymmetrical for non-Lorentzian
lines and depends on the cavity - exciton detuning. This simple
model paves the way to an heuristic derivation of eq.~\eqref{eq:formule_beta},
of practical interest to analyze the experimental behavior of an emitter~\cite{JeantetNL}. 

\begin{figure}[ht]
\includegraphics[width=8.6cm]{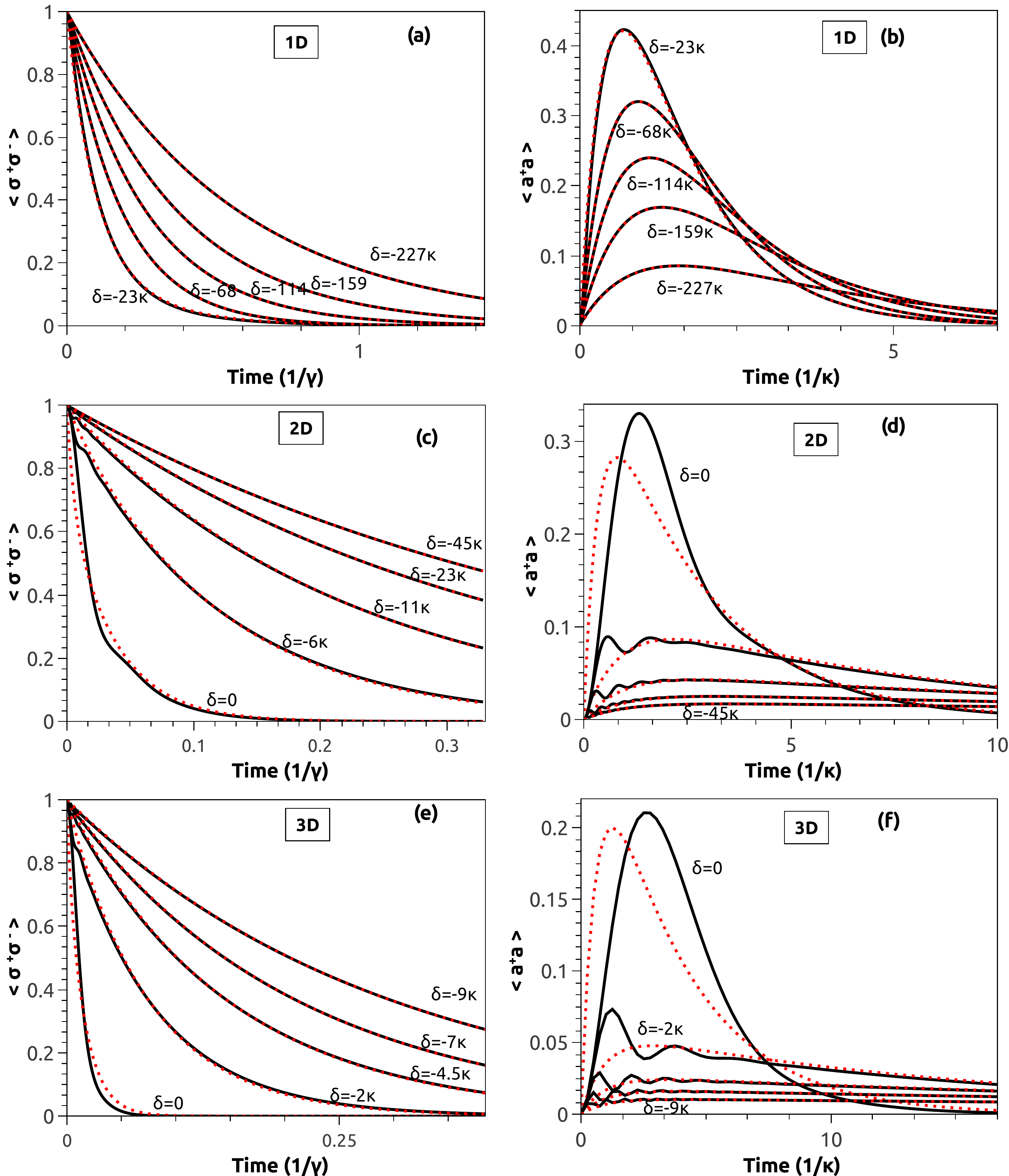} \caption{Time decay of the exciton population $\langle \hat{\sigma}^+ \hat{\sigma}^-\rangle$ (panel (a) (c) et (e) for the 1D, 2D and 3D cases resp.) and related cavity population $\langle \hat{a}^\dag \hat{a}\rangle$ (panel (b) (d) (f) resp.) for $\langle \hat{\sigma}^+\hat{\sigma}^- \rangle_{t=0}=1$ and for several detunings (expressed in units of $\kappa$). The black curves correspond to a numerical resolution of eqs.~\eqref{eq:PopEvolu1} and \eqref{eq:Population_Evolution_apres_NIBA}. The red dashed lines are obtained in the Markovian approximation from eq.~\eqref{markovdynamics}. The coupling is set to $g= \SI{250}{\micro eV}$ for the 1D case and to $g=\SI{50}{\micro eV}$ for the 2D and 3D cases. Note that the smallest detuning considered in the 1D case corresponds to the maximum efficiency (see fig.~\ref{fig:beta}). Beyond this value, one retrieves a behavior similar to that of larger detunings (i.e. as a function of $\delta$, the dynamics is symmetrical with respect to the point yielding the highest $\beta$). \label{fig:markov}}
\end{figure}

To check the consistency of the adiabatic elimination of the phonon coherence, we plot in figure~\ref{fig:markov} the population evolution computed using either the full dynamics (eqs.~\eqref{eq:PopEvolu1} and \eqref{eq:Population_Evolution_apres_NIBA}) or the Markovian approximation (eq.~\eqref{markovdynamics}). It turns out that for a 1D bath, for all detuning values and for all cavity couplings (up to $g=\SI{250}{\micro eV} $), the outcome of the Markovian approximation is almost identical to that of the full calculation. For the 2D and 3D baths, the population dynamics is well reproduce by the Markovian approximation for large detunings but becomes inaccurate for detuning smaller than the ZPL width. Note that this difference actually mainly arises from ``classical'' Rabi oscillations in the cavity-exciton sub-system. The contributions of memory effect in the phonon bath remain negligeable at this time-scale and can safely be eliminated.   

Eq.~\eqref{markovdynamics} brings a simple means to define a generalized Purcell factor as $F_{p}^{eff}(\omega_{cav})=g^{2}\tilde{S}_{\omega_{cav}}^{emi}/\gamma$.
Without phonons, when the spectrum has a Lorentzian profile, one recovers
the generalized Purcell factor obtained by Auffeves et al.~\cite{Auffeves2010}
for an emitter undergoing pure dephasing.

In the general case, the Purcell factor is thus maximum when the cavity is resonant with
the maximum of the free-space emission spectrum, which is not necessarily
located at the ZPL (\textit{e.g.} in the 1D case). Since the emission probability
$\tilde{S}^{emi}$ is a convolution
of the intrinsic spectrum with the cavity line,
it turns out that the effective Purcell factor tends to saturate when the
cavity linewidth becomes smaller than the smallest feature of the spectrum.
Any further increase in the cavity quality factor not only fails to
improve the Purcell factor, but will reduce the global single-photon
efficiency through incoherent re-absorption processes.

\subsection{Conclusion}

The interactions between an exciton, a phonon bath and a cavity are complex and an analytic solution to this problem remains
out of reach in the general case. However, a relatively simple set of assumptions, namely
the NIBA approximation and the focus on a single excitation, opens
the way to evaluate the dynamics of the system. In particular, the
outcome of the interactions between the exciton and the phonons
is encoded in the free-space emission and absorption spectra, leading
to a simple expression of the spectral efficiency of the
single-photon source.

Most quantum computing or quantum cryptography schemes require not only high purity single photon sources but also high indistinguishability~\cite{knill2001scheme,o2009photonic}.
In the weak coupling regime, this indistinguishability does not necessarily
depend on the intrinsic properties of the emitter. Indeed, the cavity
acts as the effective emitter, incoherently pumped by the pseudo two-level
system. Grange et al.~\cite{Grange1501}, have shown that two sub-regimes
exist depending on the ratio between the cavity losses and the emitter
losses (including pure dephasing). In particular, when the former
are smaller than the latter (broad emitter), the indistinguishability
can tend towards unity despite a low coupling $g$, though at the expense of brightness though.

Beyond the weak coupling regime, interesting prospects are related
to the cavity polaritons obtained when the exchanges of energy between
the cavity and the emitter become coherent. The physics of cavity
polaritons strongly depends on the dimensionality of their excitonic
part. The two-dimensional cavity polaritons have yielded Bose-Einstein
condensation~\cite{kasprzak2006bose}, as well as the generation
of twin photons~\cite{diederichs2006parametric}, while their zero-dimensional
counterparts are investigated for optical single-photon switches~\cite{volz2012ultrafast}.
More generally, their mastering is required for all optical information
processing. However, a full understanding of the role played by the
phonon bath in such a case would benefit of an interesting extension
of the work carried out here beyond the NIBA approximation.

\appendix


\section{Free-space spectra}
\label{annexefs}
\subsection{Calculation of emission and absorption spectra}

The emission and absorption spectra in free-space are described in
several publications, see for instance \cite{Mahan,Krummheuer2002a}.
As it is the starting point for further calculations, we recall the
derivation of their expression.

The damping and dephasing terms are described by Lindblad superoperators
in the master equation, given by $L_{\hat{C}}[\rho]\equiv\hat{C}\rho\hat{C}^{\dag}-\frac{1}{2}\left(\hat{C}^{\dagger}\hat{C}\rho+\rho\hat{C}^{\dagger}\hat{C}\right)$.
The exciton lifetime $\gamma$ is described by the operator $\hat{C}_{1}=\sqrt{\gamma}\hat{\sigma}^{-}$
while the pure dephasing $\gamma^{*}$, leading to a decoherent broadening
of the transition line, is described by the operator $\hat{C}_{2}=\sqrt{\gamma^{*}}\hat{\sigma}^{+}\hat{\sigma}^{-}$.
We note that the use of Lindblad superoperator is valid as we remain
in the weak coupling regime where the energy transfer is incoherent.

Another way to take into account pure dephasing is to add it as a
phenomenological dephasing factor in the response function $\chi(t)$, as done
by Krummheuer et al.~\cite{Krummheuer2002a}, which leads to the
same results.

The absorption and emission spectra are given by the Fourier Transform
of the correlation function~\cite{Mahan}: 
\begin{align}
S^{abs}(\omega)  \propto & \int_{0}^{\infty}\text{Re}\Big[e^{i\omega t}\langle\sigma^{-}(t)\sigma^{+}(0)\rangle\Big]\textrm{dt},\nonumber \\
S^{emi}(\omega)  \propto & \int_{0}^{\infty}\text{Re}\Big[e^{-i\omega t}\langle\sigma^{+}(t)\sigma^{-}(0)\rangle\Big]\textrm{dt}.
\end{align}
To compute these correlators, it is convenient to use a polaron
transformation which diagonalizes the free-space Hamiltonian: $\hat{H}_{fs}=\hat{H}_{0,fs}+\hat{H}_{X-p}$
(where $\hat{H}_{0,fs}$ is the non-interacting Hamiltonian without
the cavity term $\omega_{cav}\hat{a}^{\dag}a$). The unitary polaron
transformation $\hat{U}$ reads: 
\begin{align}
\hat{U}  \equiv & \hat{\sigma}^{+}\hat{\sigma}^{-}\otimes e^{i\Omega}+\hat{\sigma}^{-}\hat{\sigma}^{+}\otimes1\!\!1,\nonumber \\
\hat{\Omega}  \equiv & i\ \sum_{\kbo}\left(\frac{g_{\kbo}^{*}}{\omega_{\kbo}}\hat{b}_{\kbo}^{\dag}-\frac{g_{\kbo}}{\omega_{\kbo}}\hat{b}_{\kbo}\right).
\end{align}
In the following, polaron transformed operators are noted with a tilde
$\tilde{X}=\hat{U}^{\dag}\hat{X}\hat{U}$. The polaron transformed
free-space Hamiltonian is given by: 
\begin{equation}
\tilde{H}_{fs}=\hat{U}^{\dag}\hat{H}_{fs}\hat{U}=\hbar\overline{\omega}_{X}\hat{\sigma}^{+}\hat{\sigma}^{-}+\hbar\sum_{\kbo}\omega_{\kbo}\hat{b}_{\kbo}^{\dag}\hat{b}_{\kbo},
\end{equation}
The polaron shift energy is absorbed in the definition of the exciton
energy $\hbar\overline{\omega}_{X}$: 
\begin{equation}
\hbar\overline{\omega}_{X}=\hbar\omega_{X}-\frac{\hbar}{\pi}\int\frac{J(\omega)}{\omega}d\omega.
\end{equation}
In the polaron picture, the quantum evolution is given by the master
equation: 
\begin{equation}
\frac{d\tilde{\rho}}{dt}=\frac{1}{i\hbar}[\tilde{H}_{fs},\tilde{\rho}]+\sum_{i=1}^{2}L_{\tilde{C}_{i}}[\tilde{\rho}],
\end{equation}
while the Linblad operators read: $\tilde{C}_{1}=\sqrt{\gamma}\hat{\sigma}^{-}e^{i\hat{\Omega}}$
and $\tilde{C}_{2}=\hat{C}_{2}=\sqrt{\gamma^{*}}\hat{\sigma}^{+}\hat{\sigma}^{-}$
.

The expectation value of the two-time operators is given by: 
\begin{align}
\langle\hat{\sigma}^{\pm}(t)\hat{\sigma}^{\mp}(0)\rangle  = & \langle\tilde{\sigma}^{\pm}(t)\tilde{\sigma}^{\mp}(0)\rangle_{p}\nonumber \\
  = & \langle e^{\mp i\hat{\Omega}(t)}\hat{\sigma}^{\pm}(t)e^{\pm i\hat{\Omega}(0)}\hat{\sigma}^{\mp}(0)\rangle_{p},
\end{align}
where the index $p$ stands for the polaron picture \\
$\langle X\rangle=\textrm{Tr}(\rho X)=\textrm{Tr}(\tilde{\rho}\tilde{X})=\langle\tilde{X}\rangle_{p}$,
and 
\begin{equation}
\hat{\Omega}(t)=e^{it\sum_{\kbo}\omega_{\kbo}\hat{b}_{\kbo}^{\dagger}\hat{b}_{\kbo}}\ \hat{\Omega}\ e^{-it\sum_{\kbo}\omega_{\kbo}\hat{b}_{\kbo}^{\dagger}\hat{b}_{\kbo}}.\label{eq:Omega_FS}
\end{equation}
In the polaron transformed Hamiltonian, the phonon and exciton terms
commute. This implies that for a factorisable initial density matrix,
the following relation holds: 
\begin{align}
 & \langle e^{\mp i\hat{\Omega}(t)}\hat{\sigma}^{\pm}(t)e^{\pm i\hat{\Omega}(0)}\hat{\sigma}^{\mp}(0)\rangle_{p}=\nonumber \\
 & \qquad\qquad\langle e^{\mp i\hat{\Omega}(t)}e^{\pm i\hat{\Omega}(0)}\rangle_{p}\langle\hat{\sigma}^{\pm}(t)\hat{\sigma}^{\mp}(0)\rangle_{p}.
\end{align}
The excitonic part can be obtained by using the quantum regression
theorem~\cite{gardiner2004quantum}: 
\begin{equation}
\langle\hat{\sigma}^{\pm}(t)\hat{\sigma}^{\mp}(0)\rangle_{p}=e^{\pm i\overline{\omega}_{X}t-\frac{\gamma+\gamma^{*}}{2}t}\langle\hat{\sigma}^{\pm}(0)\hat{\sigma}^{\mp}(0)\rangle_{p}.
\end{equation}

By assuming that the phonon bath is at thermodynamic equilibrium,
the phonon part is given by: 
\begin{equation}
\langle e^{\mp i\hat{\Omega}(t)}e^{\pm i\hat{\Omega}(0)}\rangle_{p}=\text{Tr}\Big(\rho_{th}e^{\mp i\hat{\Omega}(t)}e^{\pm i\hat{\Omega}(0)}\Big)\equiv K(t).\label{eq:Definition_K}
\end{equation}
Where $K(t)$ is the so-called phonon kernel. The expression of $K(t)$ does not depend on the sign of $g_{\kbo}$.
Consequently $K(t)=\text{Tr}\left[\rho_{th}e^{i\hat{\Omega}(t)}e^{-i\hat{\Omega}(0)}\right]=Tr\left[\rho_{th}e^{-i\hat{\Omega}(t)}e^{i\hat{\Omega}(0)}\right]$. The expression of $K(t)$ can be computed exactly ~\cite{Leggett1987}, and its expression in the continuous limit is given in the main text (see eq. \eqref{eqKQ1Q2}). 

We finally obtain the expression of the absorption and emission spectra in freespace given in eq. \eqref{eq:fsspectra}.
We note that the integrand in  eq \eqref{eq:fsspectra} is proportional to the susceptibility $\chi(t)$
defined by Krummheuer et al.~\cite{Krummheuer2002a} in the continuous limit. 



\subsection{Phonon kernel in the long time limit}
\label{sec:Klongtime}
In the limit of long time, the $K(t)$ function can be approximated
by: 
\begin{align}
K^{(1D)}(t) & \approx  e^{-i\pi\alpha}\text{exp}[-2\pi\alpha\frac{k_{b}T}{\hbar}t],\nonumber \\
K^{(2D)}(t) & \approx  \left(\frac{1}{\omega_{c}t}\right)^{4\alpha\frac{k_{b}T}{\hbar\omega_{c}}},\nonumber \\
K^{(3D)}(t) & \approx  \text{exp}[-\sqrt{2\pi}\alpha\frac{k_{b}T}{\hbar\omega_{c}}(1-e^{-(\omega_{c}t)^{2}/8})].\label{eq:longtimeK}
\end{align}

We note that the typical decay rate of the phonon Kernel is related
to the inverse of the cut off energy $\omega_{c}$ in the case of
a super-ohmic phonon bath (2D or 3D). For an ohmic bath, the kernel
function has an exponential form, with a decay proportional to the
temperature. We also note the presence of a phase factor in the ohmic case,
which means that even in the long time limit one cannot separate the
correlator: 
\begin{equation}
\langle e^{i\hat{\Omega}(t)}e^{-i\hat{\Omega}(s)}\rangle\stackrel{t-s\rightarrow\infty}{\neq}\langle e^{-i\hat{\Omega}(s)}e^{i\hat{\Omega}(t)}\rangle.
\end{equation}
This is another way to see that the mean field theory, used in the
3D case, cannot be use for an ohmic phonon coupling.

\section{Cavity coupling}
\label{sec:cavcoupl}
Within the polaron frame, we use the interaction picture where an
operator $\tilde{X}$ reads: 
\begin{equation}
\tilde{X}_{I}(t)\equiv e^{i(\tilde{H}_{fs}+\tilde{H}_{cav})t/\hbar}\tilde{X}e^{-i(\tilde{H}_{fs}+\tilde{H}_{cav})t/\hbar},
\end{equation}
where the free cavity Hamiltonian is given by: $\tilde{H}_{cav}=\hat{U}^{\dag}\hat{H}_{cav}\hat{U}=\hbar\omega_{cav}\hat{a}^{\dagger}\hat{a}$.
The cavity losses are described with a Lindblad operator $C_{3}=\tilde{C_{3}}=\sqrt{\kappa}\hat{a}$,
where $\kappa$ is the cavity loss rate. The master equation is thus
given by: 
\begin{equation}
\frac{d}{dt}\tilde{\rho}_{I}(t)=\frac{1}{i\hbar}[\tilde{V}_{I}(t),\tilde{\rho}_{I}(t)]+\sum_{j=1}^{3}L[\tilde{C}_{I,j}(t)]\tilde{\rho}_{I}(t),\label{eq:Master_Eq_Cav}
\end{equation}
where $\tilde{V}_{I}(t)=i\hbar g\left(e^{-i\delta t}e^{i\hat{\Omega}(t)}\hat{a}^{\dag}\hat{\sigma}^{-}-e^{i\delta t}e^{-i\hat{\Omega}(t)}\hat{a}\hat{\sigma}^{+}\right)$,
and the detuning is $\delta=\omega_{X}-\omega_{cav}$ . We note that
in the interaction picture $\hat{\Omega}_{I}(t)$ is equal to the
free-space $\hat{\Omega}(t)$ defined in eq.~\eqref{eq:Omega_FS}.

The populations evolution are described by the following equations:
\begin{align}
\frac{d\langle\hat{a}^{\dag}\hat{a}\rangle}{dt}  = & -\kappa\langle\hat{a}^{\dag}\hat{a}\rangle+g\left(e^{i\delta t}\langle e^{-i\hat{\Omega}(t)}\hat{a}\hat{\sigma}^{+}\rangle+c.c.\right),\nonumber \\
\frac{d\langle\hat{\sigma}^{+}\hat{\sigma}^{-}\rangle}{dt}  = & -\gamma\langle\hat{\sigma}^{+}\hat{\sigma}^{-}\rangle-g\left(e^{i\delta t}\langle e^{-i\Omega(t)}\hat{a}\hat{\sigma}^{+}\rangle+c.c.\right).\nonumber \\
\end{align}
In order to have a closed set of equations, the evolution of the coherence
term $\langle e^{-i\Omega(t)}\hat{a}\hat{\sigma}^{+}\rangle$ is also
needed. The required trace can be conveniently split between the
system $S$ composed of an exciton and a photon (in the polaron frame) on one side 
and the phonon bath $B$ on the other side: 
\begin{equation}
\langle e^{-i\hat{\Omega}(t)}\hat{a}\hat{\sigma}^{+}\rangle=\text{Tr}_{B}\Big[\text{Tr}_{S}\big(\tilde{\rho}_{I}(t)\hat{a}\hat{\sigma}^{+}\big)e^{-i\hat{\Omega}(t)}\Big].
\end{equation}
By noting that: 
\begin{equation}
\text{Tr}_{S}\Big[\sum_{j=1}^{3}\big(L[\tilde{C}_{I,j}]\tilde{\rho}_{I}(t)\big)\ \hat{a}\hat{\sigma}^{+}\Big]=-\frac{\gamma_{all}}{2}\text{Tr}_{S}\Big[\tilde{\rho}_{I}(t)\hat{a}\hat{\sigma}^{+}\Big],
\end{equation}
where $\gamma_{all}=\gamma+\gamma^{*}+\kappa$ represents twice the
total dephasing rate.

By using the master equation given in eq. \eqref{eq:Master_Eq_Cav},
one obtains: 
\begin{align}
\frac{d}{dt}\text{Tr}_{S}\big(\tilde{\rho}_{I}(t)\hat{a}\hat{\sigma}^{+}\big)  = & \frac{1}{i\hbar}\text{Tr}_{S}\big([\tilde{V}_{I}(t),\tilde{\rho}_{I}(t)]\hat{a}\hat{\sigma}^{+}\big)\nonumber \\
   & -\frac{\gamma_{all}}{2}\text{Tr}_{S}\big(\tilde{\rho}_{I}(t)\hat{a}\hat{\sigma}^{+}\big).
\end{align}

We assume that the density matrix is diagonal at time $t=0$ (no coherence),
leading to: 
\begin{align}
\text{Tr}_{S}\big(\tilde{\rho}_{I}(t)\hat{a}\hat{\sigma}^{+}\big)  = & \frac{1}{i\hbar}\int_{0}^{t}ds\ e^{-\frac{\gamma_{all}}{2}(t-s)}\nonumber \\
   & \times\text{Tr}_{S}\Big([\tilde{V}_{I}(s),\tilde{\rho}_{I}(s)]\hat{a}\hat{\sigma}^{+}\Big).
\end{align}
Tracing the previous equation over the phonon bath yields: 
\begin{align}
   & e^{i\delta t}\langle e^{-i\hat{\Omega}(t)}\hat{a}\hat{\sigma}^{+}\rangle=g\int_{0}^{t}ds\ e^{\left(i\delta-\frac{\gamma_{all}}{2}\right)(t-s)}\nonumber \\
   & \quad\times\Big(\big\langle\hat{\sigma}^{+}\hat{\sigma}^{-}(s)e^{-i\hat{\Omega}(t)}e^{i\hat{\Omega}(s)}\big\rangle \nonumber \\
   & \qquad-\big\langle\hat{a}^{\dagger}\hat{a}(s)e^{i\hat{\Omega}(s)}e^{-i\hat{\Omega}(t)}\big\rangle\nonumber \\
   & \qquad+\big\langle\hat{a}^{\dagger}\hat{a}(s)\hat{\sigma}^{+}\hat{\sigma}^{-}(s)e^{-i\hat{\Omega}(t)}e^{i\hat{\Omega}(s)}\big\rangle\nonumber \\
   & \qquad+\big\langle\hat{a}^{\dagger}\hat{a}(s)\hat{\sigma}^{+}\hat{\sigma}^{-}(s)e^{i\hat{\Omega}(s)}e^{-i\hat{\Omega}(t)}\big\rangle\Big).\label{eq:Time_Evolution_avant_NIBA}
\end{align}
where the notation $\langle\hat{\sigma}^{+}\hat{\sigma}^{-}(s)e^{-i\hat{\Omega}(t)}e^{i\hat{\Omega}(s)}\rangle$
stands for $\text{Tr}(\rho_{I}(s)\sigma^{+}\sigma^{-}e^{-i\hat{\Omega}(t)}e^{i\hat{\Omega}(s)})$.

Let's emphasize that the time evolution given in eq. \eqref{eq:Time_Evolution_avant_NIBA}
requires no other assumption than the vanishing coherence at initial
time $t=0$.

\subsection{NIBA}

The Non-Interacting Blip Approximation (NIBA) has been introduced within a path-integral formalism by Leggett
et al. \cite{Leggett1987}. Dekker and coworkers \cite{Dekker1987}
have shown that it is equivalent to the following approximation over
the coherence term: 
\begin{align}
   & \langle\sigma^{+}\sigma^{-}(s)e^{-i\hat{\Omega}(t)}e^{i\hat{\Omega}(s)}\rangle\nonumber \\
   & \qquad\qquad\stackrel{\text{NIBA}}{\approx}\langle\sigma^{+}\sigma^{-}(s)\rangle Tr(\rho_{th}e^{-i\hat{\Omega}(t)}e^{i\hat{\Omega}(s)}),
\end{align}
where $\rho_{th}$ is the equilibrium thermal density matrix of the
bath. The validity of the NIBA in our work is ascertained in the next paragraph.
We can further simplify the expression by noting that $Tr(\rho_{th}e^{-i\hat{\Omega}(t)}e^{i\hat{\Omega}(s)})=K(t-s)$
where $K$ was introduced in eq. \eqref{eq:Definition_K} and eq. \eqref{eqKQ1Q2}.

If we restrict the system to one excitation at maximum (either a photon,
an exciton, or no excitation at all), the last two terms of eq. \eqref{eq:Time_Evolution_avant_NIBA} cancel and we obtain the population evolution given in eq. \eqref{eq:PopEvolu1} and \eqref{eq:Population_Evolution_apres_NIBA}.  




\subsection{Validity of the NIBA}
\label{sec:valNIBA}
The details of the validity of the NIBA approximation can be found in the work of Leggett et al.\cite{Leggett1987}. The condition of validity depends on the bath dimensionality. For super-ohmic coupling, and more precisely for dimension larger or equal to 2, and for all temperature, the NIBA can be applied as long as $g\ll \omega_c$. Since typical values of $\hbar \omega_c$ are in the order of $meV$, this condition is fulfilled. 

The 1D case is more problematic since non-markovian effect can become prominant.
The validity criterion for the NIBA approximation is given by ~\cite{Leggett1987}: 
\begin{equation}
\alpha k_{b}T/\hbar\gg2g(2g/\omega_{c})^{\alpha/(1-\alpha)}
\end{equation}
By noting that $(2g/\omega_{c})^{\alpha/(1-\alpha)}$ is on the order
of unity, we can simplify this condition to: 
\begin{equation}
\alpha k_{b}T/\hbar\gg2g\label{eq:inequalty_validity_niba}
\end{equation}
In the case of a $\SI{50}{\micro eV}$ cavity coupling $g$, $\hbar\omega_{c}=\SI{5}{meV}$
and $\alpha=0.25$ , the temperature has to be larger than $\SI{5}{K}$.
This condition is obtained in the case of no additional dephasing.
The left hand side of eq.~\eqref{eq:inequalty_validity_niba}
can be found in the long time behavior of $K$: 
\begin{equation}
|K(t)|\stackrel{t\rightarrow\infty}{\approx}\exp(-\frac{\alpha k_{b}T}{\hbar}2\pi\ t).
\end{equation}
From the expression of the spectra given in eq.~\eqref{eq:fsspectra},
one finds that $\frac{\alpha k_{b}T}{\hbar}\pi$ is similar to a pure
dephasing term. Consequently, the NIBA is valid when the coupling
$g$ is smaller than the temperature dependent dephasing term $2\pi\alpha k_{b}T/\hbar$,
plus the additional dephasing term $(\gamma^{*}+\gamma)/2$. In the
case of CQED, this condition corresponds to the weak coupling regime.

\subsection{Generalization to several phonon branches and optical phonons}
For the sake of simplicity we have restricted our study to the case of a coupling to a single acoustic phonon branch. In this paragraph we give the steps needed to extend the model to several phonon modes or branches including optical modes.

In the case of several branches denoted by an index $j$, we have to make the following modification in the Hamiltonian:
\begin{eqnarray}
\sum_{\kbo}  &\rightarrow &  \sum_{j,\kbo } \nonumber \\
\omega_{\kbo} &\rightarrow & \omega_{j,\kbo} \nonumber \\
 \hat{b}_{\kbo} \ ,\   \hat{b}_{\kbo}^\dagger  & \rightarrow &  \hat{b}_{j,\kbo} \ ,\    \hat{b}_{j,\kbo}^\dagger  \nonumber \\
g_\kbo   &\rightarrow & g_{j,\kbo}  
\end{eqnarray}
All the notations remain unchanged except that they refer to the $j$th phonon mode. Since the operators of different phonon modes commute, the unitary polaron transformation reads: 
\begin{align}
\hat{U}  \equiv & \big( \hat{\sigma}^{+}\hat{\sigma}^{-}\otimes  e^{i\Omega_1} \otimes  e^{i\Omega_2} \otimes ...   \big)+\big( \hat{\sigma}^{-}\hat{\sigma}^{+}\otimes1\!\!1 \otimes1\!\!1 \otimes ... \big),\nonumber \\
\hat{\Omega}_j  \equiv & i\ \sum_{\kbo}\left(\frac{g_{j,\kbo}^{*}}{\omega_{j,\kbo}}\hat{b}_{j,\kbo}^{\dag}-\frac{g_{j,\kbo}}{\omega_{j,\kbo}}\hat{b}_{j,\kbo}\right).
\end{align}
For each phonon mode $j$ one can define a phonon spectral density function $J_j(\omega) = \pi \sum_\kbo |g_{j,\kbo}|^2 \delta(\omega-\omega_{j,\kbo}) $. In the continuous limit,  it reads:
\begin{equation}
J_j(\omega)=\pi \frac{V}{(2 \pi)^s}\ DOS_j(\omega)\ |g_{j,\omega}|^2,
\end{equation}
where $s$ is the bath dimensionality, $V$ is a normalisation volume, and $DOS_j$ is the density of state of phonon mode $j$. We underline that the coupling element $g_{j,\omega}$ is a product of the bulk coupling factor with an enveloppe factor accounting for wavevector conservation in confined structures \cite{Krummheuer2002a}.

Finally, the phonon kernel $K(t)$ can be expressed as a product of single mode phonon kernels $K_j(t)$:
\begin{equation}
K(t) = \prod_j K_j(t)
\end{equation}
Most importantly, the link between the free-space and cavity-coupled properties of the emitter (as expressed by eq.~(10) of the main text) remains unchanged when considering several phonon branches.  
We have shown in the case of a single acoustic branch that the main contribution to the side-bands arises from the behaviour of $J$ near vanishing values of $\omega$. When several phonon branches are considered, the properties will be dominated by the phonon branch which brings the highest power law in the spectral density near $\omega=0$.
As for optical phonons, the phonon spectral density will be peaked around the optical phonon energy $\omega_{OP}$, giving rise to the well-known phonon replica at finite energy shift from the ZPL, but they bring negligible contribution at low energy. Thus, high-energy optical modes marginaly modify the outcome of the model when considering small spectral detunings ($\delta \ll \omega_{OP}$).

\subsection{3D case, benchmark against standard approximations:}

The approach followed in this paper to obtain the simplified populations
dynamics given in eq.~(12) is quite different from
the ones usually carried out for quantum-dots - cavity system embedded in a 3D phonon
bath. The aim of this section is to ensure that they bring similar results.

The usual starting point for a 3D phonon bath, as done with quantum dot for example by Roy et al. ~\cite{roy2011influence}, is to evaluate the difference of the bath displacement operator (called $\hat{B}$ in ~\cite{roy2011influence}, and $e^{i\hat{\Omega}}$ in the present work) regarding its thermal average value  $\langle B \rangle$. At the lowest order, if bath displacement fluctuations are omitted, the effect is to renormalize the cavity coupling constant $g$ to $\langle  B \rangle g$. This approach is very powerfull since it captures the major effect of the phonon bath through only one parameter. Unfortunately, this approach can be used only if the phonon bath dimensionality is strictly larger than two. One way to see this limitation is to consider the formal expression of $\langle B\rangle$ :
\begin{equation}
\langle B \rangle = \text{exp}\Big[ -\frac{1}{2 \pi} \int_0^\infty \frac{J(\omega)}{\omega^2} \text{coth}\left(\frac{\hbar\omega_c}{2k_b T}\right) d\omega \Big]
\end{equation}

For a dimension strictly higher than 2, the integral diverge and $\langle B \rangle =0$. At first sight, this suggests a complete cancellation of the cavity coupling, which is obviously wrong. In fact, this estimate of $\langle B \rangle$ relies on the splitting of the damping term $Q_2$ of the phonon kernel $K(t)$ (see eq.~\eqref{eqKQ1Q2}) into a constant term $Q_{2}^c$ (yielding $\langle B \rangle  = \text{exp}[-Q_2^c/(2\pi)]$) and a time varying term $Q_2^t(t)$.  This separation is possible only when the two integrals are not diverging, which is true only for a dimension strictly higher than 2. Note that the 2D case is a limit case where the divergence is logarithmic and could eventually be removed by introducing a low energy cutoff related to the physical spatial extension of the sample. Nevertheless, in 1D this approach is definitively to proscribe.

Let's now focus on a 3D bath and compare our model to existing results.
According to Roy and Hughes~\cite{roy2011influence}, a quantum dot
coupled to a cavity and a phonon bath is well approximated
by an effective Hamiltonian $H_{eff}=\hbar\omega_{x}\hat{\sigma}^{+}\hat{\sigma}^{-}+\hbar\omega_{cav}\hat{a}^{\dag}\hat{a}+i\hbar g\langle B\rangle(\hat{a}^{\dag}\hat{\sigma}^{-}-\hat{a}\hat{\sigma}^{+})$
(we restrict to a single excitation, we do not include resonant pumping). This is
the Hamiltonian of a two level system coupled to cavity where the
coupling $g$ has to be replaced by an effective coupling $g\langle B\rangle$.

In addition to the effective coupling constant, Roy et al. showed
that the effect of phonon coupling can be described by adding an effective phonon
assisted transition rate given in a Lindblad form by $L_{ph}(\rho)=\Gamma_{ph}^{\sigma^{+}a}L_{\sigma^{+}a}(\rho)+\Gamma_{ph}^{\sigma^{-}a^{\dagger}}L_{\sigma^{-}a^{\dagger}}(\rho)$.
Where the phonon mediated rates are given by: 
\begin{equation}
\Gamma_{ph}^{\sigma^{+}a/a^{\dag}\sigma^{-}}=2\langle B\rangle^{2}g^{2}\text{Re}\Big[\int_{0}^{\infty}d\tau e^{\mp i\delta\tau}\left(e^{\phi(\tau)}-1\right)\Big],
\end{equation}
where $e^{\phi(\tau)}=K(\tau)/\langle B\rangle^{2}$.

We will now show that in the weak coupling regime, the results of Roy et al., coincide with our description. 
For the Hamiltonian part, by applying an adiabatic approximation~\cite{Auffeves2010},
the cavity coupling can by replaced by an effective coupling rate
$R$ between excitation and cavity population given by: 
\begin{equation}
R=\frac{4g^{2}\langle B\rangle^{2}}{\gamma_{all}}\frac{1}{1+\left(\frac{2\delta}{\gamma_{all}}\right)^{2}},
\end{equation}

The population evolutions are then given by: 
\begin{align}
\frac{d\langle\hat{\sigma}^{+}\hat{\sigma}^{-}\rangle}{dt}  = & -(\gamma+R+\Gamma^{\sigma^{-}a^{\dag}})\langle\hat{\sigma}^{+}\hat{\sigma}^{-}\rangle\nonumber \\
& \qquad+(R+\Gamma^{\sigma^{+}a})\langle\hat{a}^{\dag}\hat{a}\rangle, \nonumber \\
\frac{d\langle\hat{a}^{\dag}\hat{a}\rangle}{dt} = & -(\kappa+R+\Gamma^{\sigma^{+}a})\langle\hat{a}^{\dag}\hat{a}\rangle\nonumber \\
& \qquad+(R+\Gamma^{\sigma^{-}a^{\dag}})\langle\hat{\sigma}^{+}\hat{\sigma}^{-}\rangle. 
\label{eq-avec-R-et-G}
\end{align}

The transition rate $R$, $\Gamma^{\hat{\sigma}^{+}\hat{a}}$ and
$\Gamma^{\hat{\sigma}^{-}\hat{a}^{\dag}}$ can be combined by noting
that $R$ can be rewritten in an integral form: 
\begin{equation}
R=2g^{2}\langle B\rangle^{2}\text{Re}\Big[\int_{0}^{\infty}d\tau e^{\pm i\delta \tau-\frac{\gamma_{all}}{2}\tau}\Big],
\end{equation}
Moreover we note that (for a 3D substrate) the function $e^{\phi(\tau)}-1$,
appearing in the phonon assisted transition rate, goes to zero with
a time of the order of $1/\omega_{c}$ (cf eq. \ref{eq:longtimeK})
which is much smaller than the inverse dephasing rate. We can add
a damping $e^{-\gamma_{all}\tau/2}$ in the definition of the phonon
mediated transition rate. We can thus combine the $R$ term and the
phonon assisted transition rate:
\begin{eqnarray}
R+\Gamma^{\sigma^{-}a^{\dag}} &\simeq & 2 g^2 \langle B\rangle^{2} \text{Re}\Big[\int_{0}^{\infty}d\tau e^{ i\delta \tau-\frac{\gamma_{all}}{2}\tau} e^{\phi(\tau)} \Big] \nonumber\\
&=& 2 g^2  \text{Re}\Big[\int_{0}^{\infty}d\tau e^{ i\delta \tau-\frac{\gamma_{all}}{2}\tau} K(\tau) \Big] \nonumber \\
&=& g^2 \tilde{S}^{emi}(\omega_{cav}).
\end{eqnarray}
Similarly we find that $R+\Gamma^{\sigma^{+}a} = g^2 \tilde{S}^{abs}(\omega_{cav})$, and the equations \eqref{eq-avec-R-et-G} are equivalent to eq. \eqref{markovdynamics}.

\begin{acknowledgments}

This work was supported by the C\textquoteright Nano IdF grant \textquotedblright ECOQ\textquotedblright{}
and the ANR grant \textquotedblleft NC2\textquotedblright . The authors thank
E. Baudin ,G. H\'etet and Ph. Roussignol for fruitful discussion.

\end{acknowledgments}

\bibliography{phonon_cavity_th}

\end{document}